\documentclass[12pt]{iopart}

\usepackage{graphicx}
\usepackage{dcolumn}
\usepackage{bm}

\usepackage[utf8]{inputenc}
\usepackage[T1]{fontenc}
\usepackage{mathptmx}
\usepackage{etoolbox}
\usepackage{xcolor}
\newcommand{\ie}{{\it i.e.}, }
\newcommand{\et}{{\it et al.} }

\begin{document}

\title[Probing NV and SiV charge state dynamics ...]{Probing NV and SiV charge state dynamics using high-voltage nanosecond pulse and photoluminescence spectral analysis}

\author{Artur Pambukhchyan}
\address{Department of Chemistry, University of Southern California, Los Angeles CA 90089, USA}

\author{Sizhe Weng}
\address{Department of Electrical Engineering, University of Southern California, Los Angeles CA 90089, USA}

\author{Indu Aravind}
\address{Department of Physics \& Astronomy, University of Southern California, Los Angeles CA 90089, USA}

\author{Stephen B. Cronin}
\address{Department of Electrical Engineering, University of Southern California, Los Angeles CA 90089, USA}

\author{Susumu Takahashi}
\ead{susumu.takahashi@usc.edu}
\address{Department of Chemistry, University of Southern California, Los Angeles CA 90089, USA}
\address{Department of Physics \& Astronomy, University of Southern California, Los Angeles CA 90089, USA}

\vspace{10pt}
\begin{indented}
\item[]July 2023
\end{indented}

\begin{abstract}
Nitrogen-vacancy (NV) and silicon-vacancy (SiV) color defects in diamond are promising systems for applications in quantum technology. 
The NV and SiV centers have multiple charge states, and their charge states have different electronic, optical and spin properties.
For the NV centers, most investigations for quantum sensing applications are targeted on the negatively charged NV (NV$^{-}$), and it is important for the NV centers to be in the NV$^{-}$ state.
However, it is known that the NV centers are converted to the neutrally charged state (NV$^{0}$) under laser excitation.
An energetically favorable charge state for the NV and SiV centers depends on their local environments. It is essential to understand and control the charge state dynamics for their quantum applications.
In this work, we discuss the charge state dynamics of NV and SiV centers under high-voltage nanosecond pulse discharges.
The NV and SiV centers coexist in the diamond crystal.
The high-voltage pulses enable manipulating the charge states efficiently.
These voltage-induced changes in charge states are probed by their photoluminescence spectral analysis.
The analysis result from the present experiment shows that the high-voltage nanosecond pulses cause shifts of the chemical potential and can convert the charge states of NV and SiV centers with the transition rates of $\sim$ MHz.
This result also indicates that the major population of the SiV centers in the sample is the doubly negatively charged state (SiV$^{2-}$), which is often overlooked because of its non-fluorescent and non-magnetic nature.
This demonstration paves a path for a method of rapid manipulation of the NV and SiV charge states in the future.
\end{abstract}

\submitto{Materials for Quantum Technology}

\section{Introduction}
Diamond is a wide bandgap semiconductor and contains many impurities and defects in the diamond lattice~\cite{Walker_1979, Ashfold2020}.
The types and concentrations of these defect centers are responsible for the color of diamond crystals.
These color centers were first studied by the gem industry, and have also become research interests for the development of high power electronics devices~\cite{KoizumiDiamondDevices}.
More recently, color defect centers in diamond have been investigated for applications of quantum technology, including quantum sensing and quantum communications.

The nitrogen-vacancy (NV) center in diamond is a fluorescent spin defect in the diamond lattice.
The NV center is promising for applications in quantum information science because of its unique properties, including the capability to polarize the spin state with optical excitation, stable photoluminescence (PL) and its long spin coherence times~\cite{Gruber1997, Jelezko2004, Epstein2005, Takahashi2008}.
NV-based quantum sensing techniques, nuclear magnetic resonance and electron spin resonance spectroscopy have been demonstrated with nanoscale sample volumes~\cite{balasubramanian08, Maze2008, Taylor2008, Abeywardana2016, Fortman2020, Fortman2021, Li2021}.
The NV center is also a great candidate for single photon sources.
Electroluminescence (EL) using a single NV center has been demonstrated~\cite{Haruyama2023, Guo2021}.
When the NV center is formed in the diamond lattice, the NV center is in one of several charge states, including negatively charged NV (NV$^-$), neutrally charged state (NV$^0$) and positively charged state NV (NV$^+$))~\cite{Gali2014}.
While most NV research is based on the NV$^-$ center,
it is important to understand and control the dynamics of the NV charge state for the formation and stabilization of the NV$^-$ state as well as applications~\cite{Fu2010, Shields2015, Jayakumar2018, Ariyaratne2018, Baier2020}.
The NV charge state can be converted under optical excitation~\cite{Aslam_2013, Shields2015, Ji2016}, electrochemical control~\cite{Karaveli2016, Hauf2011} and gate-operation of the NV centers~\cite{Grotz2012, Schreyvogel2014, Hauf2014}.

The silicon-vacancy (SiV) center is another important optically active defect in diamond.
The PL of the SiV center has been found to be particularly sensitive to the charge state of this defect. 
In particular, the negatively charged SiV$^{-}$ emits PL at 738 nm~\cite{Green2019, Muller2014, Rogers2014}, whereas the charge neutral (SiV$^0$) emits at 946 nm~\cite{D'Haenens-Johansson2011, Green2017}. 
Unlike NV emission, which is quite broad ($\sim$ 200 nm) at room temperature due to a wide array of phonon sidebands, the SiV feature is spectrally narrow (4 nm at room temperature). 
These spectrally narrow photons have a high degree of indistinguishability and are better able to couple to photonic crystal cavities and waveguides making them promising candidates for solid-state-based quantum technologies.
Single SiV defects have been detected in single crystal diamond~\cite{Neu_2011}.
The SiV defect provides a unique molecular structure with the interstitial silicon atom being placed between two unoccupied carbon sites~\cite{Goss1996}. 
This results in an inversion symmetry ($D_{3d}$) of the split-vacancy configuration, which is subsequently responsible for the high spectral stability as it protects the optical transition from local electric field fluctuations~\cite{Hepp2014}. 
Studies of the electronic structure of the defect have revealed an $S = 1$ ground state for the neutral SiV~\cite{Rogers2014} and a $S = 1/2$ ground state for negative configuration~\cite{D'Haenens-Johansson2011, Dhomkar2018, Green2019}. 
In addition, a double negatively charged state (SiV$^{2-}$) is also expected to be stable in diamond, however, SiV$^{2-}$ has no optical transition and accessible spin levels, and it may be difficult to be detected~\cite{Gali2013, Breeze2020}.
It is known that the charge state of SiV$^{-}$ centers can be converted under optical illumination~\cite{Dhomkar2018, Nicolas2019}. 
However, the nature of the charge state and the mechanisms of the conversion are not known well.

Charge state control of the NV and SiV centers by applied electric fields may be useful for some applications in quantum devices.
Doi \et reported electrical charge state manipulation of NV centers in diamond using a p-i-n heterostructure device~\cite{Doi2014}.  Here, the charge state was modulated via a current driven effect (rather than a voltage driven effect), in which NV$^0$ states were induced by current injection. In contrast, the more recent work of Weng \et uses high voltage pulses to overcome the large Schottky barrier at the copper/diamond interface, injecting charge, and resulting in SiV$^{-}$ charge state conversion~\cite{Weng2021}. In addition, this high voltage approach does not require heterostructure fabrication and is particularly advantageous considering the relatively large contact resistances associated with this wide band gap semiconductor. 

In this work, we probe the charge state dynamics of NV and SiV centers in diamond with the application of high voltage nanosecond pulses and the analysis of their PL spectra.
We employ a type-IIa diamond, which contains both NV and SiV centers.
Recent work by Weng \et shows the efficient increase of the SiV population under high voltage nanosecond pulses~\cite{Weng2021}. 
However, the efficiency of the charge conversion and the mechanism of the charge state dynamics have yet to be fully understood.
Here, we discuss the charge conversion of NV centers under high voltage nanosecond pulses and characterize their conversion rates quantitatively.
The high voltage pulses are typically 10-20 ns long and are $\sim$ 5 kV in magnitude.
The high voltage pulses are repetitively applied to convert the NV charge states.
The repetition rate of the high-voltage pulse is varied from 1 Hz to 1 kHz to modulate the charge conversion dynamics.
The PL measurement of both NV and SiV centers is performed by acquiring the PL signal with application of a continuous-wave (CW) laser excitation and repetitive high-voltage pulses.
We analyze the PL spectra to extract the populations of the negatively charged state of NV centers ([NV$^{-}$]) and neutrally charged states of NV centers ([NV$^0$]). 
We also analyze the change in the population of the SiV$^{-}$ states.
The result shows that $\sim$13 \% of the NV$^-$ population is converted to NV$^0$ with the application of the high-voltage pulses with a repetition rate of 1 kHz.
The population of the SiV$^{-}$ is also increased by $\sim$20 times.
This study paves a way for the method development of fast manipulation of the NV and SiV charge states.

%
\section{Experiment}
\begin{figure}[ht]
\begin{center}
\includegraphics[width = 16 cm]{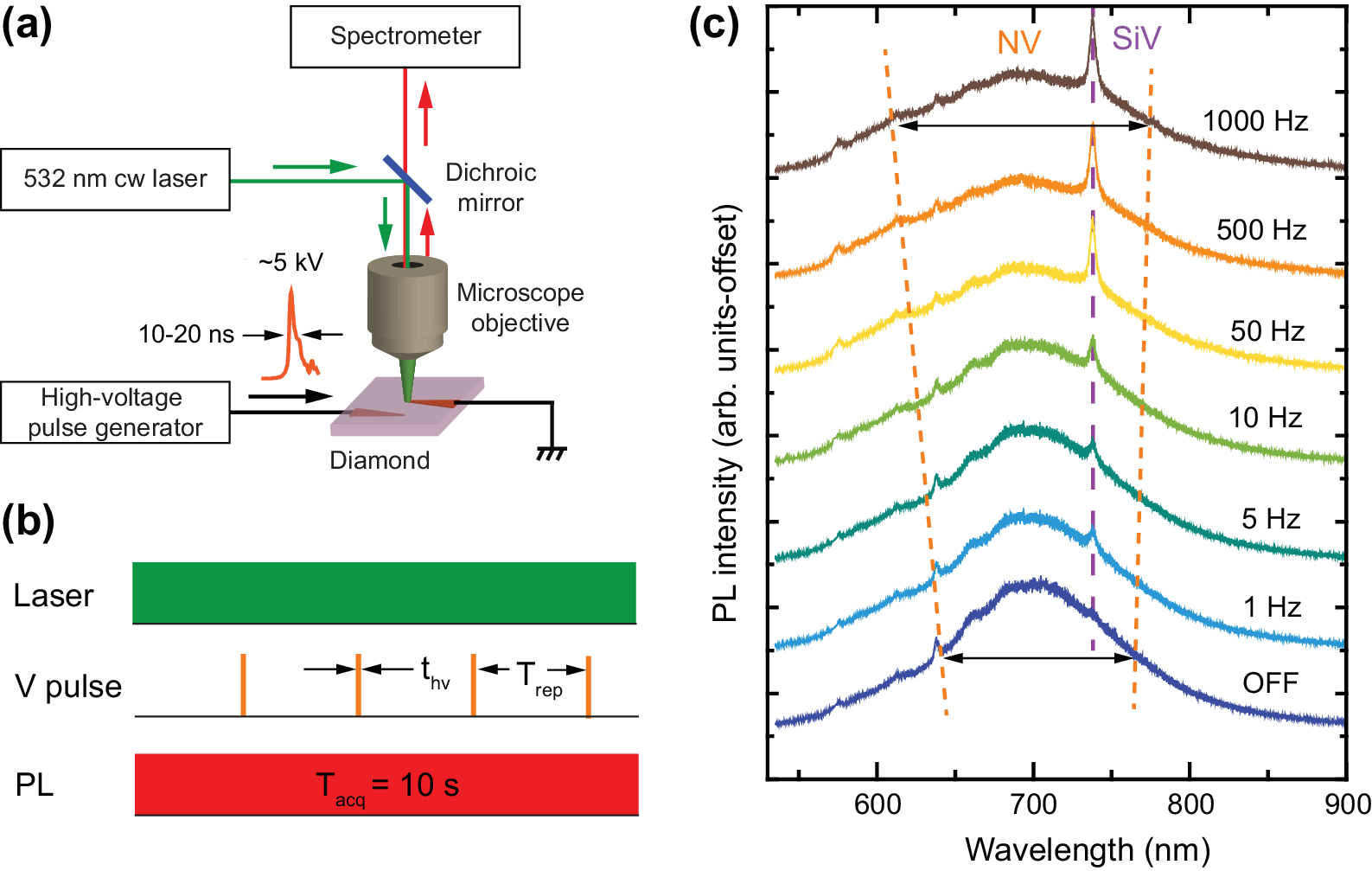}    
\end{center}
\caption{
Overview of the PL measurement.
(a) Schematic diagram of the experimental setup showing a typical lineshape of the high-voltage pulse.
(b) Diagram representing the experimental sequence, in which the high-voltage pulse has the duration of $t_{hv}$, the repetition period is $T_{rep}$, and the acquisition time ($T_{acq}$) is 10 s in the present case.
(c) PL spectra plotted as a function of wavelength. The spectra were collected with various repetition rates. The full-widths at half-maximum of NV PL spectra are indicated by orange dashed lines. 
The peak position of the SiV PL emission is indicated by the purple dashed line.
}
\label{fig:Setup}
\end{figure}
In this experiment, polycrystalline type IIa CVD diamond (Element Six, Ltd., Thermal Management Grade, TM220) was studied.
The size of the diamond crystal is 10$\times$10$\times$0.5 mm$^3$. No processing was applied on the diamond sample. 
The experimental setup, as depicted in Fig.~\ref{fig:Setup}(a), involved attaching strips of copper tape that were manually cut to create sharp tips with a radius of 20 $\mu$m on both the top and bottom of the diamond sample. 
These sharp electrode tips served to enhance the local electric field across the diamond sample when the high voltage is applied. The separation between the two electrodes is 0.5 mm in the lateral direction~\cite{Weng2021}.
A continuous wave (CW) laser with a wavelength of 532 nm was focused on the top surface of the diamond using a microscope objective lens with a numerical aperture of 0.6 and a working distance of 4 mm (Olympus ULCPlanFLN). 
The bottom electrode was connected to a pulse generator (SSPG-20X, Transient Plasma Systems, Inc.) capable of producing high-voltage pulses with a rise time of 5-10 ns ($dV/dt=10^{12}$ V/s) at a pulse repetition rate ranging from 1 to 1000 Hz. The top electrode was grounded.
Figure 1(a) displays the typical output characteristics of the nanosecond pulses applied to the bottom electrode. PL spectra of the diamond sample were captured from the diamond sample's top surface, near the tip of the top electrode, using an \textit{inVia} micro-spectrometer (Renishaw, Inc.). 

As shown in Fig.~\ref{fig:Setup}(b), the PL measurement was performed under CW laser excitation, PL acquisition and repetitive excitation of high voltage nanosecond pulses.
The power of the excitation laser was 37 $\mu$W.
The PL acquisition time ($T_{acq}$) was set to 10 s for the PL measurement.
The PL measurements were performed with and without the application of high voltage nanosecond pulses.
The repetition rate of high-voltage nanosecond pulses was varied between 1 - 1000 Hz (the corresponding repetition period ($T_{rep}$) is from 1 ms to 1 s).
As shown in Fig.~\ref{fig:Setup}(a), the high-voltage pulses typically have a duration of 10-20 ns and a peak voltage of up to $\sim$ 5 kV.

Figure~\ref{fig:Setup}(c) shows the PL spectra of NV centers with and without the application of the high-voltage pulses.
The repetition rate of the high-voltage pulses was varied in 1-1000 Hz.
As shown in Fig.~\ref{fig:Setup}(c), the NV spectra become broader by applying the high-voltage pulses at higher repetition rates.
This broadening effect is much larger than that due to Stark shifts~\cite{Tamarat2006}.
In addition, as reported previously~\cite{Weng2021}, the high-voltage pulses also increases the PL intensity of the SiV$^-$ emission observed at 738 nm.

\section{Analysis methods}
\begin{figure}[ht]
\begin{center}
\includegraphics[width = 9 cm]{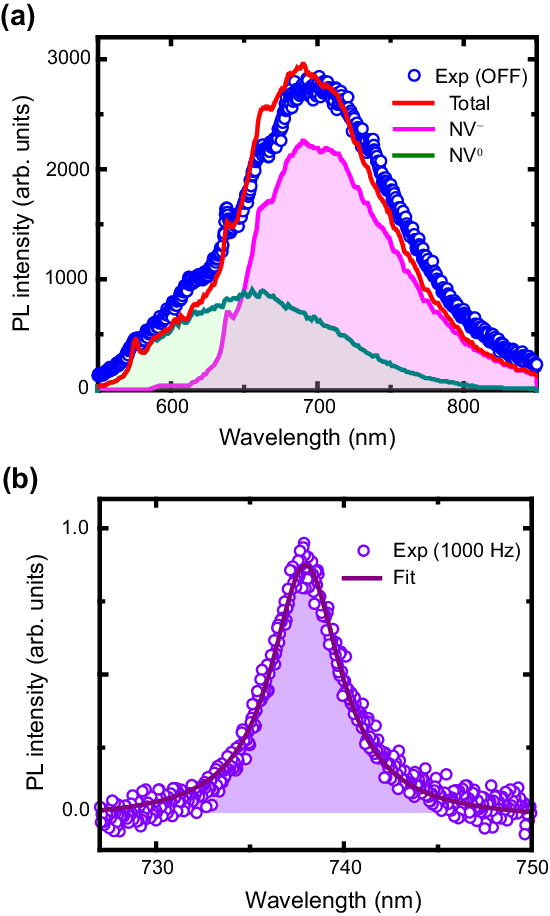}    
\end{center}
\caption{
Population determination method using diamond PL spectra. 
(a) Depiction of the PL spectral decomposition method to extract the population of NV$^{-}$ and NV$^0$ centers.
The PL spectrum shown here was taken without the high-voltage pulse excitation.
(b) Background-subtracted PL spectrum used for the determination of the SiV$^{-}$ population.
The PL spectrum was taken with a repetition rate of 1000 Hz.
}
\label{fig:PL}
\end{figure}
\subsection{Determination of [NV$^-$] and [NV$^0$] using PL decomposition analysis}
Here, we discuss a method to estimate the populations of NV$^-$ and NV$^0$ centers using the PL decomposition analysis~\cite{Alsid2019}.
The population of the NV$^+$ state is not considered here.
In the decomposition analysis, we consider an experimentally observed PL spectrum (I($\lambda$)) to be the sum of the NV$^-$ and NV$^0$ PL spectra.
Namely, I($\lambda$) is expressed by the following equation,
\begin{equation}
        I(\lambda) = a_1 I^-(\lambda) + a_2 I^0(\lambda)
\label{Eq:PLInt}
\end{equation}
where $I^-(\lambda)$ and $I^0(\lambda)$ are the normalized PL spectrum of NV$^-$ and NV$^0$ centers, respectively.
$I^-(\lambda)$ and $I^0(\lambda)$ may be the data reported previously.
In the present case, we used $I^-(\lambda)$ and $I^0(\lambda)$ extracted from Ref.~\cite{Jeske2017}.
In addition, $a_1$ and $a_2$ in Eq.~\ref{Eq:PLInt}, representing the contributions from NV$^-$ and NV$^0$ centers on the PL spectrum, are fitting parameters.
$a_1$ and $a_2$ are proportional to the populations of NV$^-$ and NV$^0$ centers as well as their PL emission efficiency.
Therefore, by taking into account the emission efficiency ratio between NV$^-$ and NV$^0$ centers, the populations of NV$^-$ and NV$^0$ can be expressed by,
\begin{equation}
\eqalign{ [NV^-] = \frac{a_1}{a_1 + \kappa_{532} a_2}, \cr
[NV^0] = 1 - [NV^-] = \frac{\kappa_{532} a_2}{a_1 + \kappa_{532} a_2},}
\label{Eq:PopNVm}
\end{equation}
where $\kappa_{532}$ is the PL emission efficiency ratio between NV$^-$ and NV$^0$ centers under 532 nm wavelength excitation, and was estimated to be 2.5 previously~\cite{Alsid2019}.
Fig.~\ref{fig:PL}(a) shows the analysis result on the PL data without high voltage pulse.
As can be seen, the fit agrees well with the experimental data.
With the fit and Eqs.~\ref{Eq:PLInt} and ~\ref{Eq:PopNVm}, we obtained [NV$^-$] $=  46 \pm 1 \%$  and [NV$^0$] $=  54 \pm 1 \%$.

\subsection{Estimate of SiV$^{-}$ population change}
The population of charged SiV$^{-}$ centers under high-voltage pulses is estimated by analyzing the intensity of the PL spectrum.
As shown in Fig.~\ref{fig:Setup}(c), the SiV$^{-}$ emission was observed at a wavelength of 738 nm.
Since the PL peak area is proportional to the SiV$^{-}$ population ([SiV$^{-}$]), the population change due to the application of high voltage pulses can be estimated by analyzing the peak areas.  
Figure~\ref{fig:PL}(b) shows the SiV PL spectrum with background correction.
In the correction, the background from the NV PL signal is removed by dividing the background from the signal data. 
Then, the peak area ($I_{SiV-}$) is obtained by fitting the data with a Lorentzian function, as shown in Fig.~\ref{fig:PL}(b).
In the present case, we analyzed the PL data taken with a repetition rate of 1000 Hz and obtained $I_{SiV}=6.2 \pm 0.1$ 
Using the value of $I_{SiV}=6.2$ for 1000 Hz, the normalized peak area ($\tilde{I}_{SiV-}$) at each repetition rate was calculated, namely, $\tilde{I}_{SiV-} = I_{SiV}/6.2$.

\begin{figure}[ht]
\begin{center}
\includegraphics[width = 14 cm]{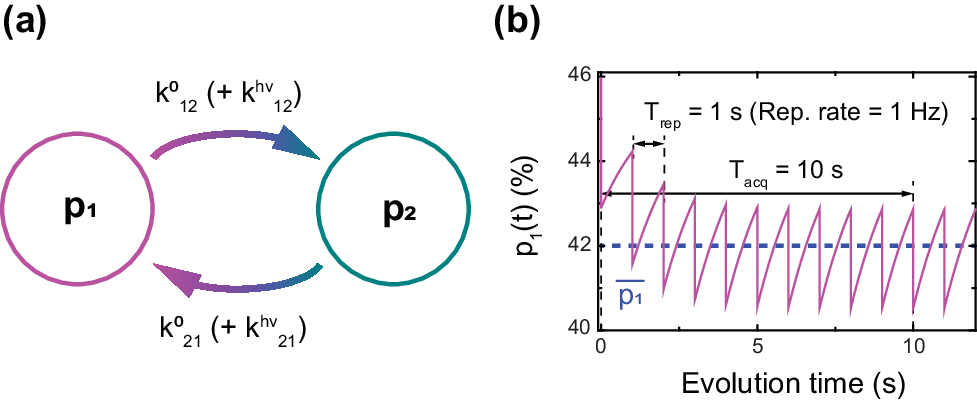}    
\caption{
Depiction of the simulation method for NV and SiV charge state dynamics.
(a) Two-state model to understand diamond PL spectra with charge conversion dynamics.
(b) Time evolution of the $p_1$ population. 
The calculation was performed with the repetition rate of 1 Hz ($T_{rep} = 1$ s), $k^{0}_{12}=0.3$ $s^{-1}$, $k^{0}_{21}= (46/54)k^{0}_{12}$, $k^{hv}_{12}=18.5\times 10^6$ $s^{-1}$ and $k^{hv}_{21}=9.2\times 10^6$ $s^{-1}$. 
$\bar{p_1}=42$\% in the present case.}
\label{fig:PopSim}
\end{center}
\end{figure}
\subsection{Simulation of two-charge state dynamics} \label{Sect:sim}
In order to understand the repetition rate dependence of the PL spectra of NV and SiV centers, we model their charge state dynamics as a two-state system. 
Figure~\ref{fig:PopSim}(b) shows the model, consisting of the population proportions of State 1 and 2 (denoted as $p_1$ and $p_2$).
$k_{12}$ and $k_{21}$ are the transition rates from State 1 to 2 and from State 2 to 1, respectively.
The time evolution of $p_1$ and $p_2$ are given by the following system rate equations,
\begin{equation}
\eqalign{ \frac{dp_{1}(t)}{dt} = -k_{12}p_{1}(t) + k_{21}p_2(t), \cr
         \frac{dp_2(t)}{dt} = k_{12}p_{1}(t) - k_{21}p_2(t),}       
        \label{Eq:PopNVTrans}
\end{equation}
where
\begin{equation*}
k_{12 (21)}=\cases{k^0_{12 (21)} + k^{hv}_{12 (21)} & during $t_{hv}$-period\\
k^0_{12 (21)} & during $T_{rep}$-period\\}
\end{equation*}
$k^0_{12}$ and $k^0_{21}$ are the transition rates of $p_1$ $\rightarrow$ $p_2$ and $p_2$ $\rightarrow$ $p_1$ processes without the application of high-voltage pulse.
$k^{hv}_{12}$ and $k^{hv}_{21}$ are the transition rates induced by the high-voltage pulse.
Using the measurement sequence shown in Fig.~\ref{fig:Setup}(b) and Eq.~\ref{Eq:PopNVTrans}, the time evolution of $p_1(t)$ and $p_2(t) = 1 - p_1(t)$ is calculated. 
In the calculation, we consider that the high-voltage pulse is 10 ns long (\ie $t_{hv}=10$ ns) and $k^{hv}_{12}$ and $k^{hv}_{21}$ are constant during $t_{hv}$.
Figure~\ref{fig:PopSim}(b) shows the result of $p_1(t)$.
As shown in Fig.~\ref{fig:PopSim}(b), $p_1(t)$ decreases rapidly by the nanosecond voltage pulse, and then $p_1(t)$ relaxes back after the nanosecond pulse.
However, $p_1(t)$ does not reach the initial state before the next nanosecond pulse because both $k^{0}_{12}$ and $k^0_{21}$ rates are slower than the repetition rate of $1/T_{rep} = 1$ Hz.
Then, when the second nanosecond pulse is applied, $p_1(t)$ decreases further. In a longer time scale, $p_1(t)$ is oscillatory and the center of the oscillations decreases asymptotically.
From the time evolution result with $T_{acq}=10$ s, the mean values of $p_1(t)$ and $p_2(t)$ ($\overline{p_1}$ and $\overline{p_2}$) are calculated to compare with [NV$^-$] and [NV$^0$] as well as the change of the normalized PL area of SiV$^{-}$ ($\tilde{I}_{SiV-}$).
In the next section, we discuss the population analyses for the NV and SiV centers.
The NV population is analyzed by considering the NV$^{-}$ population as $p_{1}(t)$ and the NV$^{0}$ population as $p_{2}(t)$.
In the SiV analysis, SiV$^{-}$ population is considered as $p_{2}(t)$.

%
%
\section{Results and Discussion}
\begin{figure}[ht]
\begin{center}
\includegraphics[width = 16 cm]{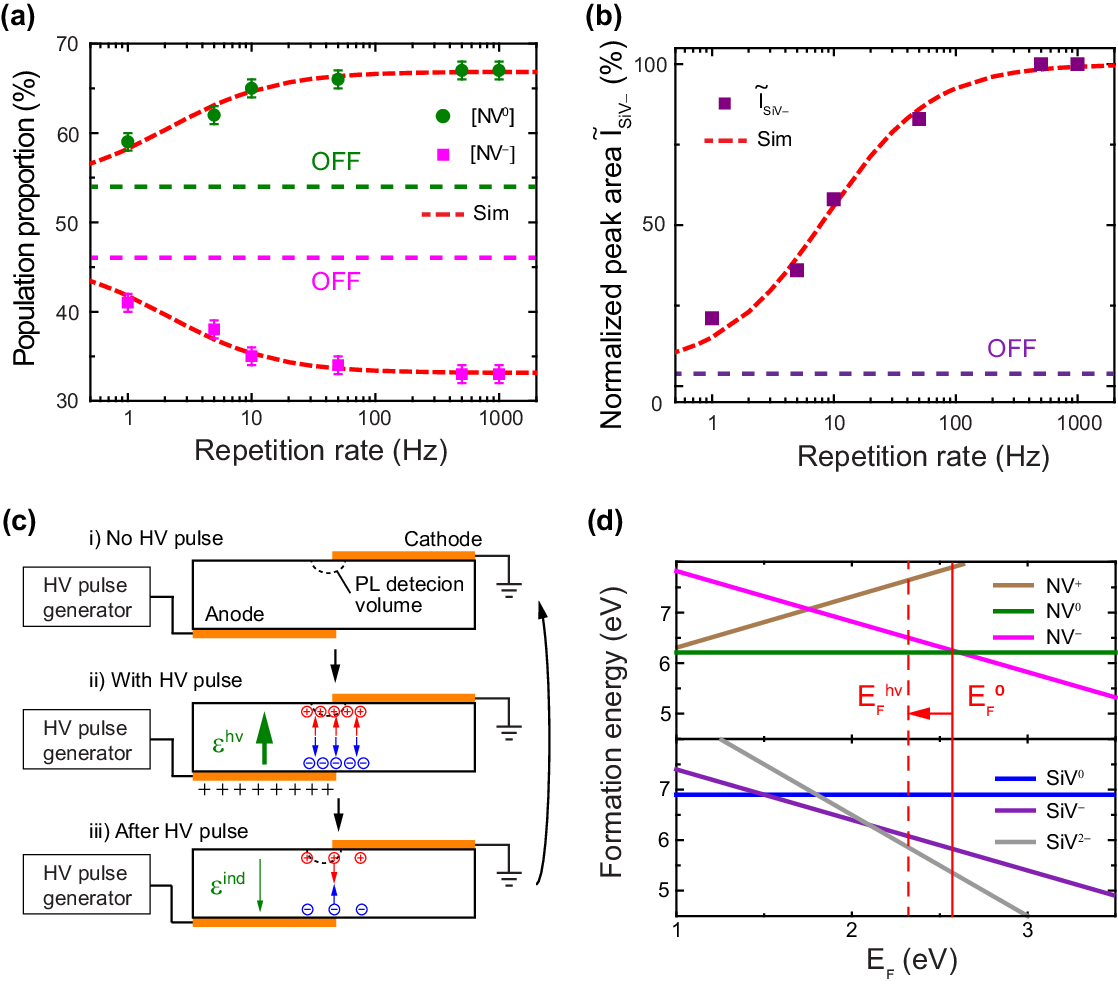}   
\caption{
Analysis of the NV and SiV charge state modulation.
(a) Populations of NV$^-$ and NV$^0$ plotted as a function of the repetition rate of the high-voltage pulses. The y-axis shows the population proportions of NV$^{-}$ and NV$^0$. The x-axis represents the repetition rate in Hz (1/s).  
The data without the high-voltage pulse (indicated as OFF) is shown by dashed lines.
The error bars represent 95 \% confidence intervals.
The error bars are calculated using the obtained uncertainties of $a_1$ and $a_2$ and the error propagation.
(b) The normalized peak areas of SiV$^{-}$ PL spectra plotted as a function of the repetition rates.
(c) Schematic diagram depicting the charge dynamics with the application of the high-voltage pulses.
(d) The formation energy of the NV and SiV charge states plotted as a function of the chemical potential ($E_F$). See Ref.~\cite{Gali2013, Gali2013-2} for details about the charge formation energy levels.
}
\label{fig:Pop}
\end{center}
\end{figure}
\begin{table}[ht]
    \centering
\begin{tabular}{||c c c c c c c||}
\hline
Rep. rate (Hz) & [NV$^-$] & [NV$^0$] & $2\sigma_{NV-}$ & $2\sigma_{NV0}$ & $\tilde{I}_{SiV-}$ & $2\sigma_{SiV-}$\\ 
\hline\hline
OFF & 46 & 54 & 1 & 1 & 5 & 1\\
1 & 41 & 59 & 1 & 1 & 21 & 1\\
5 & 38 & 62 & 1 & 1 & 36 & 1\\
10 & 35 & 65 & 1 & 1 & 58 & 1\\
50 & 34 & 66 & 1 & 1 & 83 & 1\\
500 & 33 & 67 & 1 & 1 & 100 & 1\\
1000 & 33 & 67 & 1 & 1 & 100 & 1\\
\hline\hline
\end{tabular}
    \caption{Obtained [NV$^-$], [NV$^0$] and the normalized SiV PL intensity ($\tilde{I}_{SiV-} = I_{SiV-}/6.2$) values and their uncertainties. 2$\sigma$ represents 95 \% confidence intervals.}
    \label{tab:NVb1b2}
\end{table}
\subsection{NV Centers}
Figure~\ref{fig:Pop}(a) and Table~\ref{tab:NVb1b2} show the NV$^-$ and NV$^0$ populations proportions ([NV$^-$] and [NV$^0$], respectively) as a function of the repetition rate of the applied high-voltage pulses. 
The populations were obtained using the PL decomposition analysis with Eqs.~\ref{Eq:PLInt} and \ref{Eq:PopNVm}.
As can be seen in Fig.~\ref{fig:Pop}, when the repetition rate increases, [NV$^-$] decreases while [NV$^0$] increases.
In the present experiment, [NV$^-$] ([NV$^0$]) was changed from 46 \% (54 \%) to 33 \% (67 \%) when the repetition rate is varied from 0 Hz (corresponding to no voltage pulse applied) to 1000 Hz.
This population changes caused the PL broadening shown in Fig.~\ref{fig:PL}.
As shown in Fig.~\ref{fig:Pop},
the changes of the populations is more pronounced in the repetition range of 1-100 Hz, and the populations are nearly constant above the repetition of rate of 100 Hz.
To understand the charge state dynamics, the result was fitted with the simulation described in the previous section. 
We consider $p_1$ and $p_2$ in Fig.~\ref{fig:PopSim} to be NV$^{-}$ and NV$^0$ states, respectively.
In the calculation of Eq.~\ref{Eq:PopNVTrans}, we set the initial values of [NV$^-$] and [NV$^0$] to be the values from the OFF result, namely, 46 \% and 54 \%, respectively.
In addition, the OFF high-voltage pulse result sets $k^{0}_{21}/k^{0}_{12} = 46/54$.
Thus, the model has three parameters, namely, $k^{0}_{12}$,  $k^{hv}_{12}$ and $k^{hv}_{21}$.
Figure~\ref{fig:Pop}(a) shows the fit result of the experiment and model.
As one can see, the model explains the experimental data very well.
From the fit, we obtained $k^0_{12} = 0.3\pm0.1$ Hz and $k^{hv}_{12} = (18.5 \pm 0.1) \times 10^6$ Hz and $k^{hv}_{21} = (9.2 \pm 0.3) \times 10^6$ Hz.
Therefore, the result showed that the NV charge conversion rates by the high-voltage nanosecond pulse is in the range of MHz.

\subsection{SiV Centers}
Figure~\ref{fig:Pop}(b) shows the normalized PL intensity of the SiV$^{-}$ centers plotted as a function of the repetition rate.
The PL intensity without the application of high voltage pulses is also indicated by the purple dashed line.
The PL intensity is proportional to the SiV$^{-}$ population.
As can be seen, the PL intensity increased from OFF to 1000 Hz by $\sim$20 times, indicating a significant increase of the SiV$^{-}$ population.
Moreover, we perform the PL measurement of SiV$^0$ at 946 nm (not shown), and no PL signal was observed. However, it is possible that phonon broadening effects on SiV$^0$ PL hamper the observation at room temperature since most of the previous PL measurements of SiV$^0$ have been done at low temperatures~\cite{D'Haenens-Johansson2011, Green2017, Rose2018}.
Therefore, in Fig.~\ref{fig:Pop}(b), we showed the SiV$^{-}$ population normalized by the value measured at the repetition rate of 1000 Hz ($\tilde{I}_{SiV-}$) instead of the SiV$^{-}$ portion in all SiV charge state populations since we have no information of the populations of other SiV charge states such as SiV$^0$ and SiV$^{2-}$.
Similarly to the NV centers (see Fig.~\ref{fig:Pop}(a)), the PL intensity changes are pronounced in the range of 1-100 Hz, indicating the time scale of the SiV charge state dynamics is similar to that of the NV centers. 
Finally, we determine the charge conversion rates using the model shown in Sect.~\ref{Sect:sim}. 
In the analysis, we set $k^0_{12}/k^0_{21} = 5/95$ and $k^0_{21}=0$ because of the initial and final states of $\tilde{I}_{SiV-}$.
By fitting with Eq.~\ref{Eq:PopNVTrans}, we obtained $k^0_{12} = 0.033\pm0.007$ Hz and $k^{hv}_{12} = (8.58 \pm 0.01) \times 10^6$ Hz.

\subsection{Charge dynamics with high-voltage nanosecond pulse}
As can be seen in many semiconductors, band bending occurs when an electric field created by a high voltage pulse ($\epsilon^{hv}$) is applied to the diamond crystal.
Then, the band bending causes electrons moves to the high voltage side and creates the unbalance of the charge density over the sample volume, with which the anode size has higher electron density and the cathode side has lower electron density (see Fig.~\ref{fig:Pop}(c)). 
In addition, the charge unbalance induces an electric field oppose to the high-voltage pulse ($\epsilon^{ind}$ in Fig.~\ref{fig:Pop}(c)). 
Once the high-voltage pulse is off, electrons diffuse over the sample volume to cancel the charge separation.
With the use of the high-voltage nanosecond pulse, electrons get a high kinetic energy and the charge movement process happens in short time while the charge diffusion is a slow process.
The modulation of the electron density also changes the chemical potential. 
Since the present experiment measures the PL signals from the small detection volume on the cathode side, the NV and SiV centers in the detection volume experience a lower chemical potential than that with no high-voltage pulse. 
Figure~\ref{fig:Pop} shows the formation energies of NV and SiV charge states obtained from Ref.~\cite{Gali2013, Gali2013-2}.
As can be seen, NV$^{-}$ charge state is stable when the chemical potential is $\sim$2.6 eV above the valence band energy (the chemical potential ($E_{F}$) is defined from the valence band energy here).
When $E_{F}$ is below $\sim$2.6 eV, NV$^0$ is more stable.
For SiV centers, the SiV$^{2-}$ charge state is energetically favorable when $E_{F}$ is above $\sim$2.2 eV. 
When $E_{F}$ is between 2.2 and 1.5 eV, the SiV$^{-}$ is stable. 
Then, SiV$^0$ is favorable, when $E_{F}$ is below 1.5 eV.

In the present experiment, we observed that the [NV$^-$] and [NV$^0$] values are similar when no high-voltage pulse is applied. 
This indicates the chemical potential with no high-voltage pulse ($E_F^0$) is $\sim$2.6 eV.
Then, if $E_F$ decreases with the application of the high-voltage pulses ($E_F^{hv}$), [NV$^-$] will decrease and [NV$^0$] will increase.
This is consistent with the observation as shown in Fig.~\ref{fig:Pop}(a). 
For the SiV center, since $E_F^{hv}$ gets closer to the SiV$^{-}$ and SiV$^{2-}$ crossing energy of 2.2 eV, SiV$^{-}$ will also increase.
Again, this is consistent with the observation on the SiV$^{-}$ population change.
Therefore, the SiV result strongly supports that the SiV$^{-}$ centers are converted from the SiV$^{2-}$ centers.

%
%
\section{Summary}
In summary, we investigate the charge state dynamics of NV and SiV centers under the application of high voltage nanosecond pulses and analyze their PL spectra.
The results show that the application of the high-voltage pulses converts the NV populations from NV$^{-}$ to NV$^0$.
Moreover, a significant increase of the SiV$^{-}$ populations was observed with the application of the high-voltage pulses. 
We show that the observed charge state dynamics can be explained by a simple two state model and transition rates of MHz were extracted.
The changes of the populations of both NV and SiV centers are consistent with the lowering of the chemical potential of the diamond by the high voltage pulses.
The result also supports that the charge state of SiV centers in the diamond sample is non-fluorescent and non-magnetic SiV$^{2-}$ centers.
The experimental method can be further developed by combining with time-domain measurements to study the charge dynamics at faster time scales and multi-charge state processes as well as to demonstrate the manipulation of NV and SiV charge states with a fast electric field pulse.

\ack
This research was supported by the National Science Foundation (NSF) Award No. ECCS-2204667 (S.W. and S.B.C.) and the Army Research Office (ARO) Award No. W911NF2210284 (I.A. and S.B.C.).
A.P. and S.T. also thanks the supports from NSF (ECCS-2204667 and CHE-2004252 with partial co-funding from the Quantum Information Science program in the Division of Physics), the USC Anton B. Burg Foundation, and the Searle scholars program.

\section*{Data availability statement}
The data that support the findings of this study are available from the corresponding author upon reasonable request.

\section*{References}

\end{document}